\documentclass{emulateapj}

\usepackage{rotate}
\usepackage{longtable}
\usepackage{color}

\shorttitle{AGN Activity in Compact Star-Forming Galaxies at $z\sim2$}
\shortauthors{Kocevski et al.}

\begin{document}

\title{CANDELS: Elevated Black Hole Growth in the Progenitors of Compact Quiescent Galaxies at $z\sim2$}

\author{Dale D.~Kocevski\altaffilmark{1}, 
Guillermo Barro\altaffilmark{2}, 
S.M.~Faber\altaffilmark{3}, 
Avishai Dekel\altaffilmark{4},
Rachel S.~Somerville\altaffilmark{5,6},
Joshua A.~Young\altaffilmark{1},
Christina C.~Williams\altaffilmark{7},
Daniel H. McIntosh\altaffilmark{8},
Antonis Georgakakis\altaffilmark{9},
Guenther Hasinger,\altaffilmark{10},
Kirpal Nandra\altaffilmark{9},
Francesca Civano\altaffilmark{11},
David M.~Alexander\altaffilmark{12},
Omar Almaini\altaffilmark{13},
Christopher J.~Conselice\altaffilmark{13},
Jennifer L.~Donley\altaffilmark{14},
Harry C.~Ferguson\altaffilmark{15},
Mauro Giavalisco\altaffilmark{16},
Norman A.~Grogin\altaffilmark{15},
Nimish Hathi\altaffilmark{15},
Matthew Hawkins\altaffilmark{1},
Anton M.~Koekemoer\altaffilmark{15},
David C.~Koo\altaffilmark{3},
Elizabeth J.~McGrath\altaffilmark{1},
Bahram Mobasher\altaffilmark{17},
Pablo G.~P\'erez Gonz\'alez\altaffilmark{18},
Janine Pforr\altaffilmark{19},
Joel R.~Primack\altaffilmark{20},
Paola Santini\altaffilmark{21},
Mauro Stefanon\altaffilmark{22,23},
Jonathan R.~Trump\altaffilmark{24},
Arjen van der Wel\altaffilmark{25},
Stijn Wuyts\altaffilmark{26},
Haojing Yan\altaffilmark{23}}

\affil{$^1$ Department of Physics and Astronomy, Colby College, Waterville, ME 04961; dale.kocevski@colby.edu\\ 
	   $^2$ Department of Physics, University of the Pacific, Stockton, CA, USA\\
	   $^3$ UCO/Lick Observatory, Department of Astronomy and Astrophysics, University of California, Santa Cruz, CA, USA\\
	   $^4$ Center for Astrophysics and Planetary Science, Racah Institute of Physics, The Hebrew University, Jerusalem, Israel\\
	   $^5$ Department of Physics and Astronomy, Rutgers, The State University of New Jersey, Piscataway, NJ, USA\\
	   $^6$ Center for Computational Astrophysics, Flatiron Institute, New York, NY, USA\\
	   $^7$ Steward Observatory, University of Arizona, Tucson, AZ, USA\\
	   $^8$ Department of Physics \& Astronomy , University of Missouri-Kansas City, Kansas City, MO, USA\\
	   $^9$ Max-Planck-Institut f\"ur extraterrestrische Physik, Garching, Germany\\
	   $^{10}$ Institute for Astronomy, University of Hawaii, Honolulu, HI, USA\\
	   $^{11}$ Department of Physics, Durham University, Durham, UK\\
	   $^{12}$ Harvard-Smithsonian Center for Astrophysics, Cambridge, MA, USA\\
	   $^{13}$ School of Physics \& Astronomy, University of Nottingham, Nottingham, UK\\
	   $^{14}$ Los Alamos National Laboratory, Los Alamos, NM, USA\\
	   $^{15}$ Space Telescope Science Institute, Baltimore, MD, USA\\
	   $^{16}$ Department of Astronomy, University of Massachusetts, Amherst, MA, USA\\
	   $^{17}$ Department of Physics and Astronomy, University of California, Riverside, CA, USA\\
	   $^{18}$ Departamento de Astrof\'isica, Universidad Complutense de Madrid, Madrid, Spain\\
	   $^{19}$ European Space Research and Technology Centre, European Space Agency, Noordwijk, The Netherlands \\
	   $^{20}$ Department of Physics, University of California, Santa Cruz, CA, USA\\
	   $^{21}$ Osservatorio Astronomico di Roma, Monte Porzio Catone (RM), Italy\\
	   $^{22}$ Leiden Observatory, Leiden University, Leiden, Netherlands\\
	   $^{23}$ Department of Physics and Astronomy, University of Missouri, Columbia, MO, USA\\
	   $^{24}$ Department of Physics, University of Connecticut, Storrs, CT, USA\\
	   $^{25}$ Max-Planck Institut f\"ur Astronomie, Heidelberg, Germany\\   
	   $^{26}$ Department of Physics, University of Bath, Bath, UK }

\begin{abstract}

We examine the fraction of massive ($M_{*}>10^{10} M_{\odot}$), compact star-forming galaxies (cSFGs) that host an active galactic nucleus (AGN) at $z\sim2$.  These cSFGs are likely the direct progenitors of the compact quiescent galaxies observed at this epoch, which are the first population of passive galaxies to appear in large numbers in the early Universe.  We identify cSFGs that host an AGN using a combination of \emph{Hubble} WFC3 imaging and \emph{Chandra} X-ray observations in four fields: the \emph{Chandra} Deep Fields, the Extended Groth Strip, and the UKIDSS Ultra Deep Survey field.  We find that $39.2^{+3.9}_{-3.6}$\% (65/166) of cSFGs at $1.4<z<3.0$ host an X-ray detected AGN.  This fraction is 3.2 times higher than the incidence of AGN in extended star-forming galaxies with similar masses at these redshifts.  This difference is significant at the $6.2\sigma$ level.  Our results are consistent with models in which cSFGs are formed through a dissipative contraction that triggers a compact starburst and concurrent growth of the central black hole.   We also discuss our findings in the context of cosmological galaxy evolution simulations that require feedback energy to rapidly quench cSFGs.  We show that the AGN fraction peaks precisely where energy injection is needed to reproduce the decline in the number density of cSFGs with redshift.  Our results suggest that the first abundant population of massive, quenched galaxies emerged directly following a phase of elevated supermassive black hole growth and further hints at a possible connection between AGN and the rapid quenching of star formation in these galaxies.

\end{abstract}

\keywords{galaxies: active --- galaxies: evolution --- X-rays: galaxies}

\section{Introduction}

One of the key goals of galaxy evolution studies is understanding how massive, passively evolving galaxies observed in the local universe obtained their present-day properties.  In particular, a substantial amount of work has gone into determining when these quenched galaxies formed the bulk of their stars, how they grew their central supermassive black holes (SMBH) and by what process was their star formation activity shut down.  While the stellar populations of these galaxies suggest an early formation epoch ($z>2$; e.g., McCarthy et al.~2004; Renzini et al.~2006; Cimatti et al.~2008)), studies have revealed significant growth in their stellar mass and number density since $z\sim1$ (Bell et al.~2004; Faber et al.~2007; Brown et al.~2007), indicating that both the growth of individual galaxies through merging and the continued quenching of massive, star-forming galaxies is required to reproduce the build-up of the red sequence over time.

More recently, deep near-infrared surveys with \emph{Hubble} have extended these studies to $1\!<\!z\!<\!3$, the epoch where roughly half of all present-day stellar mass is formed (e.g., Dickinson et al.~2003; Fontana et al.~2006) and quiescent galaxies start to appear in large numbers for the first time.  The quenched fraction among massive galaxies ($M\!>\!10^{11}$ $M_{\odot}$) increases from as low as $\sim7\%$ at $z\sim3$ to $90\%$ at $z=1$ (Whitaker et al.~2010; Marchesini et al.~2010; Bell et al.~2012; Muzzin et al.~2013; Ilbert et al.~2013; Tomczak et al.~2014; Straatman et al.~2015) and the stellar mass density of quiescent galaxies increases sharply between $z=2$ and $z=1$ (Arnouts et al.~2007; Ilbert et al.~2010; Brammer et al.~2011).  This first generation of quiescent galaxies are the likely progenitors of the most massive, early-type galaxies found locally (e.g., Hopkins et al.~2009a, van Dokkum et al.~2014) and studying the processes that give rise to this population is central to understanding the origins of their local counterparts.

A key characteristic of quiescent galaxies at $z\sim2$ is their compact size (e.g., Daddi et al.~2005; Trujillo et al.~2006; van Dokkum et al.~2008; Williams et al.~2010; Cassata et al.~2011).  Passive galaxies with stellar masses of $M\sim10^{11}$ $M_{\odot}$ are $\sim$4 times smaller at $z=2$ than they are at $z=0$ (van der Wel et al.~2014) and two orders of magnitude more dense than their local counterparts (van Dokkum et al.~2008; Bezanson et al.~2009).  These compact quiescent galaxies (cQGs) dominate the early-type population at this redshift, making up 90\% of massive, quenched galaxies at $z=2-3$ (Cassata et al.~2013). 

Studies have found that the number density of cQGs steadily increases between $z=3$ and $z=1.5$ and then quickly declines at $z<1$ (Cassata et al.~2013; van der Wel et al.~2014; van Dokkum et al.~2015), suggesting they must experience a significant amount of size growth at later times.  Many mechanisms have been proposed to achieve this growth, with the prevailing theory being dry (gas-poor), non-dissipative mergers that add mass to the outskirts of galaxies without initiating new rounds of star formation (Naab et al.~2009; Hopkins et al.~2010; Oser et al. 2010; Porter et al. 2014).  It is through this process that cQGs are thought to eventually become giant ellipticals on the high-mass end of the red sequence at low redshifts (e.g., van Dokkum et al.~2014).

There is currently much debate as to how massive, compact galaxies formed in the early universe.  Several theories have been proposed, many of which rely on high gas fractions and highly dissipative processes to achieve the extreme stellar densities observed in cQGs.  This includes gas-rich mergers (Mihos \& Hernquist 1996; Hopkins et al.~2009b; Wuyts et al.~2010; Wellons et al.~2015) and the compaction of extended star-forming galaxies due to violent disk instabilities (VDI; Dekel et al.~2009; Dekel \& Burkert 2014; Zolotov et al.~2015; Tacchella et al.~2016a).  In both cases, tidal torques act to funnel gas (and stellar clumps in the case of VDI) to small radii, which triggers a nuclear starburst and ultimately results in a compact remnant.  Alternatively, the dense cores of cQGs may have formed in situ at even higher redshifts, when all galaxies were denser, and the resulting galaxies remained compact until $z\sim2$ (Wellons et al.~2015; Lilly \& Carollo 2016; Williams et al.~2017).

Since all of these formation scenarios require large supplies of cold gas and intense star formation on nuclear scales, it stands to reason that compact star-forming galaxies (cSFGs) should also be detected.  Indeed, several studies have now reported finding star-forming galaxies with effective radii, velocity dispersions, and stellar densities comparable to that of the cQGs population (Barro et al.~2013, 2014, Stefanon et al.~2013; Patel et al.~2013; Williams et al.~2014; Nelson et al.~2014).  These cSFGs are heavily obscured by dust (Barro et al.~2014; Nelson et al.~2014), have extreme star formation rates of 200-700 $M_{\odot}$ yr$^{-1}$ (Barro et al.~2016), and appear to have ubiquitous AGN activity (Rangel et al.~2014).  Barro et al.~(2013) showed that the number density of cSFGs since $z\sim3$ declines at a rate that matches the increase in density of cQGs assuming a quenching timescale of $0.3-0.8$ Gyr.  Based on this observation, they proposed that the cSFGs found at $z=2-3$ are the direct progenitors of the cQGs that are seen to build up at this same epoch.

Given the structural similarities between the two populations, all that is needed to convert cSFGs into their quiescent counterparts is the truncation of their star-formation activity.  This naturally raises the question of how this quenching is achieved. Although cSFGs have short gas consumption timescales ($t\sim230$ Myr; Barro et al.~2016; see also Spilker et al.~2016; Popping et al.~2017; Tadaki et al.~2017) due to their high rate of star formation, simulations suggest that this alone is not enough to produce the level of inactivity observed in cQGs without additional mechanisms to prevent renewed star formation in the future (Zolotov et al.~2015; Tacchella et al.~2016a, 2016b).  It has been proposed that these mechanisms are related to the rapid build-up of the central bulge that happens during this phase, as high surface mass densities have long been linked to quiescence (e.g., Kauffman et al.~2003; Brinchmann et al.~2004; Franx et al.~2008; Bell et al.~2011; Cheung et al.~2012; van Dokkum et al.~2014; Woo et al.~2015; Whitaker et al.~2017).  This may be due to a form of morphological quenching, where the high central density stabilizes the surrounding gas against gravitational collapse (Martig et al.~2009; Genzel et al.~2014).  Alternatively, it may be due to the growth of the central SMBH and feedback from the resulting AGN (Springel et al.~2005; Hopkins et al.~2006; Croton et al.~2006).  Along with energy injection from supernova, this feedback can drive outflows that help to deplete the galaxy's cold gas supply and/or provide heating that prevents gas cooling and future star formation.

\begin{figure*}[t]
\epsscale{1.15}
\plotone{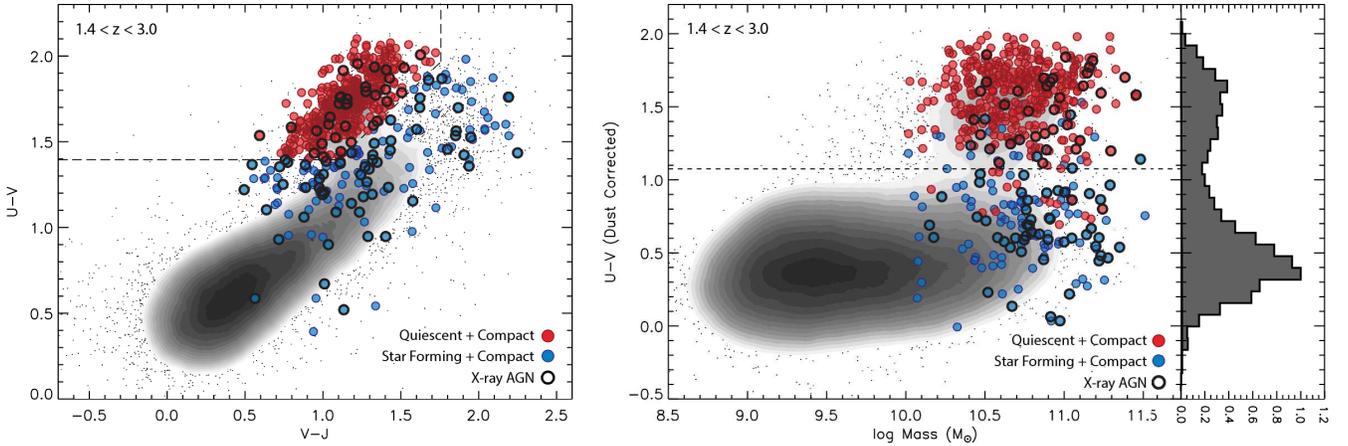}
\caption{(\emph{left}) UVJ color diagram for galaxies in the redshift range $1.4<z<3.0$ in the CANDELS fields.  The grey contours/points show the color distribution of the general galaxy population.  The dashed line denotes the dividing line used to separate quiescent and star-forming galaxies.  Massive (log $M_{*} > 10^{10} M_{\odot}$) compact quiescent galaxies are shown in red, while massive compact star-forming galaxies are shown blue.  Black circles indicate galaxies that host an X-ray bright AGN. (\emph{right}) Stellar mass plotted against dust-corrected $U-V$ color for the same galaxies shown in the left panel.  Our parent sample shows a color bimodality at $U-V=1.08$.  This single color cut is effective at separating the star-forming and quiescent galaxies selected using our $UVJ$ color cut.
  \label{fig-UVJ}}  
\end{figure*}

In this paper, we examine the prevalence of X-ray selected AGN in cSFGs at $z\sim2$ in order to shed light on the connection between this phase of galaxy evolution and the growth of SMBHs at high redshift, as well as the role that AGN feedback may play in quenching the star-formation activity of these galaxies and ultimately giving rise to the cQG population.  As the progenitors of today's giant ellipticals and their massive central SMBHs, cSFGs can provide an important window into how the AGN-galaxy connection is established and maintained at high redshifts.

We present our analysis as follows. \S2 describes the optical, near-infrared and X-ray data used for this study, while \S3 details our methodology for selecting cSFGs that host AGN and how we test for AGN contamination of our morphology measurements.  In \S4 we present our primary results and discuss their implications in the context of cosmological galaxy evolution simulations that require feedback energy to quench the cSFG population.  Finally, we summarize our findings in \S5.  Throughout this paper we assume the cosmological parameters $(\Omega_{\rm M}, \Omega_{\rm \Lambda}, h) = (0.27,0.73, 0.71)$.

\section{Data Description}

Our parent sample of massive galaxies is drawn from HST/WFC3 F160W (\emph{H}-band) selected catalogs in four of the five CANDELS fields (Grogin et al 2011; Koekemoer et al. 2011).  This includes the Great Observatories Origins Deep Survey (GOODS; Giavalisco et al.~2004) North and South fields, the UKIDSS Ultra Deep Survey (UDS; Lawrence et al.~2007, Cirasuolo et al.~2007), and the Extended Groth Strip (EGS, Davis et al.~2007).  
Point source depths vary among the CANDELS fields from $H=27$ in the wide fields to $H=27.7$ in the deep fields (see Grogin et al.~2011).  Multi-wavelength photometry ($U$-band to 8$\mu$m) was measured in each field using the TFIT routine (Laidler et al.~2006) as described in detail in Guo et al.~(2013), Galametz et al.~(2013), Stefanon et al.~(2017), and Barro et al.~(2017, in press) for the GOODS-S, UDS, EGS, and GOODS-N fields, respectively.
Photometric redshifts were computed in each field using the method described in Dahlen et al.~(2013) and resulted in typical errors of $\Delta z/(1+z)=3\%$ at $z>1.5$.  Stellar masses were computed as described in Mobasher et al.~(2015) and Santini et al.~(2015).  Rest-frame photometry was derived by fitting templates to the observed-frame spectral energy distributions (SEDs) using the EAZY code (Brammer et al.~2008), as described in Kocevski et al.~(2017, in prep.).  Visual extinction values, $A_{V}$, were derived using FAST (Kriek et al.~(2009) assuming a Chabrier (2003) initial mass function, solar metallicity, exponentially declining star formation histories, and the Calzetti et al.~(2000) dust extinction law (see Wuyts et al.~2011 for additional details).  Galaxy morphologies and sizes were measured from the HST/WFC3 \emph{H}-band images using GALFIT (Peng et al.~2002) as described in van der Wel et al.~(2014).  This includes Sersic indicies and effective (half-light) radii.

X-ray detections in all fields except the UDS come from publicly available \emph{Chandra} point source catalogs.  In GOODS South and North, we make use of the 4 Ms and 2 Ms point source catalogs of Xue et al.~(2011) and Xue et al.~(2016), respectively, while for EGS, we use the 800 ks source catalog presented in Nandra et al.~(2015).   In UDS, we use a source catalog from the X-UDS survey (PI. G.~Hasinger; Kocevski et al.~2017).  These observations consist of 25 \emph{Chandra}/ACIS-I pointings mosaiced to achieve $\sim600$ ks depth in the area of UDS imaged by CANDELS.

\section{Sample Selection}

In this study, we aim to determine the fraction of massive, compact star-forming galaxies (cSFGs) at $z\sim2$ that host an X-ray bright AGN, relative to their extended star-forming counterparts (eSFGs).  We start with an initial sample of 12,975  $H$-band selected galaxies over our four CANDELS fields in the redshift range $1.4<z<3.0$ and which are brighter than $H=24.5$, the magnitude limit down to which galaxy sizes can be accurately determined (van der Wel et al.~2012).  We next apply a stellar mass cut of $M_{*}>10^{10} M_{\odot}$ and we exclude bright point sources by ensuring all sources have a SExtractor (Bertin \& Arnouts 1996) stellarity index of {\tt CLASS\_STAR} $<0.9$.  This results in a sample of 3199 massive galaxies, which are hereafter referred to as our parent sample.

Next we split our parent sample into star-forming and quiescent galaxies using a $U-V$ and $V-J$ color cut based on that of Williams et al.~(2009).  In particular, star-forming galaxies were selected as those that satisfy the following two criteria:

\begin{equation}
U-V < 0.85\times(V-J)+0.46 
\end{equation}
\begin{equation}
U-V < 1.4
\end{equation}

\noindent The rest-frame $UVJ$ colors of CANDELS galaxies in our redshift window are shown in Figure 1.  The dashed line denotes our adopted boundary between star-forming and quiescent galaxies, which is based on the color bimodality observed in our initial sample of 12,975 galaxies at this redshift.  

It should be noted that minor changes to our adopted $UVJ$ boundary do not significantly affect our results.  In fact, we find that a simple cut on $U-V$ color corrected for dust using a Calzetti et al.~(2000) attenuation law works equally well in selecting star-forming systems because the galaxies in our parent sample show a clear color bimodality at $(U-V)_{\rm corr} = 1.08$.  Therefore, we highlight this single color threshold as our dividing line when plotting surface mass density against rest-frame color throughout the remainder of this paper.

\begin{figure}[t]
\epsscale{1.1}
\plotone{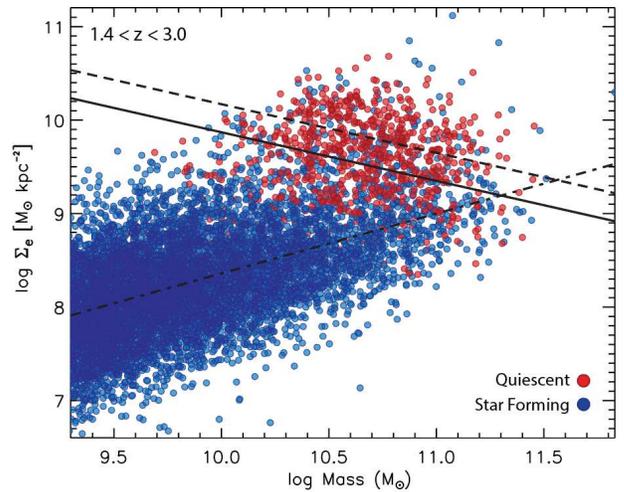}
\caption{Surface mass density ($\Sigma_{e}=0.5M/\pi r^{2}_{e}$) measured within the effective radius versus stellar mass for galaxies at $1.4<z<3.0$.  The blue and red circles show star-forming and quiescent galaxies selected using our UVJ color criteria shown in Figure \ref{fig-UVJ}.  The dashed and dashed-dotted lines denote the best-fit log $\Sigma$ - log $M_{*}$ relationship of Barro et al.~(2017) for quiescent and star-forming galaxies, respectively.  The solid black line marks our surface density threshold above which galaxies are considered compact for their given mass.  
  \label{fig-MassSigma}}  
\end{figure}

Next, following Barro et al.~(2017), compact galaxies are selected using a mass-dependent surface density threshold.  Figure \ref{fig-MassSigma} shows the surface mass density measured at the effective radius, $\Sigma_{e} = 0.5M_{*}/\pi r^{2}_{e}$, versus mass for galaxies in our redshift window of $1.4<z<3.0$.  Galaxies are separated into star-forming and quiescent systems based on our $UVJ$ color selection.  These two populations follow well-defined size-mass relationships of the form ${\rm log} ~r_{e} \propto a\,{\rm log}\,M$ (e.g., Newman et al.~2012; van der Wel et al.~2014), which, expressed in terms of $\Sigma_{e}$, take the form:

\begin{equation}
{\rm log}~\Sigma_{e} = \alpha \Big[ {\rm log}~\Big(\frac{M_{*}}{M}\Big) - 10.5 \Big] + {\rm log~A}
\end{equation}

\noindent Here $\alpha$ is related to the slope of the size-mass relationship, $a$, as $\alpha = 1-2a$, and A is the overall normalization.  The best-fit $\Sigma_{e}$-mass relationship for star-forming and quiescent galaxies at $z\sim2.2$, as determined by Barro et al.~(2017), is shown in Figure \ref{fig-MassSigma} as the dashed and dot-dashed lines, respectively.  

As demonstrated in van der Wel et al.~(2014), quiescent galaxies $z=2-3$ are $\sim$4 times more compact than their star-forming counterparts at a given mass.  In order to identify the likely star-forming progenitors of these quiescent galaxies, we define cSFGs as systems that have surface mass densities similar to that of the quiescent population.  More specifically, cSFGs are selected as galaxies that satisfy our star-forming $UVJ$ color cut and those that fall within 0.3 dex of the $\Sigma_{e}$-mass relationship for quiescent galaxies. 
For this purpose, we use the best-fit parameters from Barro et al.~(2017), namely $\alpha=-0.52\pm0.14$ and log A$=9.91\pm0.07$.  This structural criteria is shown as the solid black line in Figure \ref{fig-MassSigma}; all star-forming galaxies that lie above this line are considered compact for their given mass and redshift.  A value of 0.3 dex is roughly $1.2\times$ the intrinsic scatter in the quiescent $\Sigma_{e}$-mass relationship at our target redshift.  It should be noted that our results are not sensitive to changes of up to 50\% in this adopted threshold; i.e., our findings on the relative AGN fraction in cSFGs versus eSFGs is statistically unchanged using thresholds ranging from 0.15-0.45 dex.

To identify cSFGs that host AGN, optical counterparts to X-ray sources in each field except the UDS were taken from the literature.  In GOODS-N and GOODS-S, we adopt the \emph{H}-band counterparts provided in Xue et al.~(2016) and Hsu et al.~(2014), while in EGS we use the counterparts identified in Nandra et al.~(2015).  In the UDS, we matched the X-UDS source catalog to the CANDELS $H$-band catalog of Galametz et al.~(2013) using the maximum likelihood technique described in Sutherland \& Saunders (1992) and more recently implemented by Civano et al.~(2012).  In short, the method gauges the likelihood that a $H$-band source is matched to an X-ray source by comparing the probability of finding a genuine counterpart with the positional offset and magnitude of the optical candidate relative to that of finding a similar object by chance.  Likelihood ratios were calculated for all galaxies within $5^{\prime\prime}$ of an X-ray source, taking into account the positional uncertainty of the X-ray centroid and the magnitude of the possible counterpart galaxy.  A likelihood threshold is set which maximizes both the completeness and reliability of the crossmatches (see Civano et al.~2012 for details) and optical matches with likelihood ratios above this threshold are deemed secure.

In each field, X-ray luminosities in the soft (0.5-2 keV), hard (2-8 keV) and full (0.5-8 keV) bands are then computed from the observed fluxes in each band using the best available CANDELS redshift (which are a combination of ground-based spectroscopic and photometric redshifts) and $K$-corrected assuming a power-law spectrum with a spectral slope of $\Gamma=1.4$.  Sources with X-ray luminosities in excess of $10^{42}$ erg s$^{-1}$ in any band are then flagged as AGN since the X-ray emission from high-mass X-ray binaries in star-forming galaxies rarely exceeds this luminosity  (Alexander et al.~2005).

In the following analysis, we combine the AGN detected in all four fields into a single sample despite the different X-ray flux limits of the \emph{Chandra} datasets.  This is because our primary objective is to determine the \emph{relative} difference in the AGN fraction between compact and extended star-forming galaxies.  Since these galaxies are uniformly distributed among our target fields, differences in survey depth will be reflected in the AGN fractions of both populations.  In total, 323 galaxies from our parent sample were identified as hosting an X-ray AGN.

\subsection{Testing for AGN Contamination}

Since finding cSFGs relies on accurate mass, color, and structural measurements, contamination by non-stellar light from a central AGN is a potential concern.  The most severe contamination would be expected from luminous, unabsorbed (type I) AGN.  Fortunately, given the limited survey area covered by our four fields (0.206 deg$^2$), we do not expect a large number of such AGN in our sample.  For example, Hsu et al.~(2014) modeled the spectral energy distribution (SED) of X-ray sources in GOODS-S using galaxy+AGN hybrid templates and found only five luminous (log $L_{\rm X} > 44$), type I AGN in our redshift window whose nuclear emission dominated the light of their host galaxy.  Extrapolating this result, we expect to find $\sim22$ such sources in our four fields.  If all of these type I AGN survive our initial selection criteria (i.e., cuts on mass, magnitude and stellarity), we expect at most 6.8\% (22/323) of our X-ray sources will be severely contaminated by non-stellar light. 

To further mitigate the affects of AGN contamination, we have excluded from our analysis all unresolved AGN hosts, as well as extended hosts that show point-like emission at their centers.  Point-like emission was identified using a combination of surface brightness profile fitting (see below), visual inspection of the host morphology, and two-dimensional Galfit modeling.  A total of 13 AGN hosts identified as cSFGs or eSFGs at $1.4<z<3.0$ were excluded based on these tests.

\begin{figure}[t]
\epsscale{1.18}
\plotone{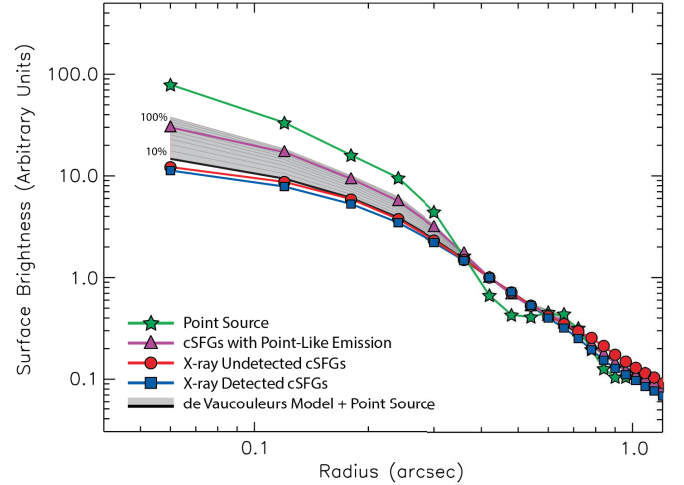}
\caption{Stacked $H$-band (F160W) radial surface brightness profiles for cSFGs that host an AGN (blue squares) and those that do not (red circles).  Also shown is the stacked profile of cSFGs identified as suffering from nuclear contamination (magenta triangles) and the profile for point source emission (green stars).  For comparison, the solid black line is the surface brightness expected for a pure de Vaucouleurs profile, to which we have added point-like emission ranging from 10\% to 100\% of the model galaxy's total integrated light.  We find that the stacked profiles of the cSFGs which host AGN and those that do not are in good agreement, implying minimal contamination.  
  \label{fig-radialprof}}  
\end{figure}

To look for nuclear contamination in our remaining sample, we have stacked the surface brightness profiles of the X-ray detected and non-detected cSFGs after excluding contaminated sources.  These stacked profiles are shown in Figure \ref{fig-radialprof}.    Also shown is the surface brightness profile of a pure de Vaucouleurs model, to which we have added point-like emission of varying strength, ranging from 10\% to 100\% of the model galaxy's total integrated light.  Even moderate nuclear emission is easily visible as a steepening of the surface brightness profile.  This is clearly evident in the stacked profile of the cSFGs which have been excluded from our analysis because they suffer from point-like AGN contamination.  The stacked profile of these contaminated sources is also shown in Figure \ref{fig-radialprof} for comparison.  Most importantly, we find that the stacked profiles of the cSFGs which host AGN and those that do not are in excellent agreement, suggesting that, on average, less than 10\% of the rest-frame optical light from the X-ray detected cSFGs originates from an unresolved nuclear component.

\begin{figure*}[t]
\epsscale{1.15}
\plotone{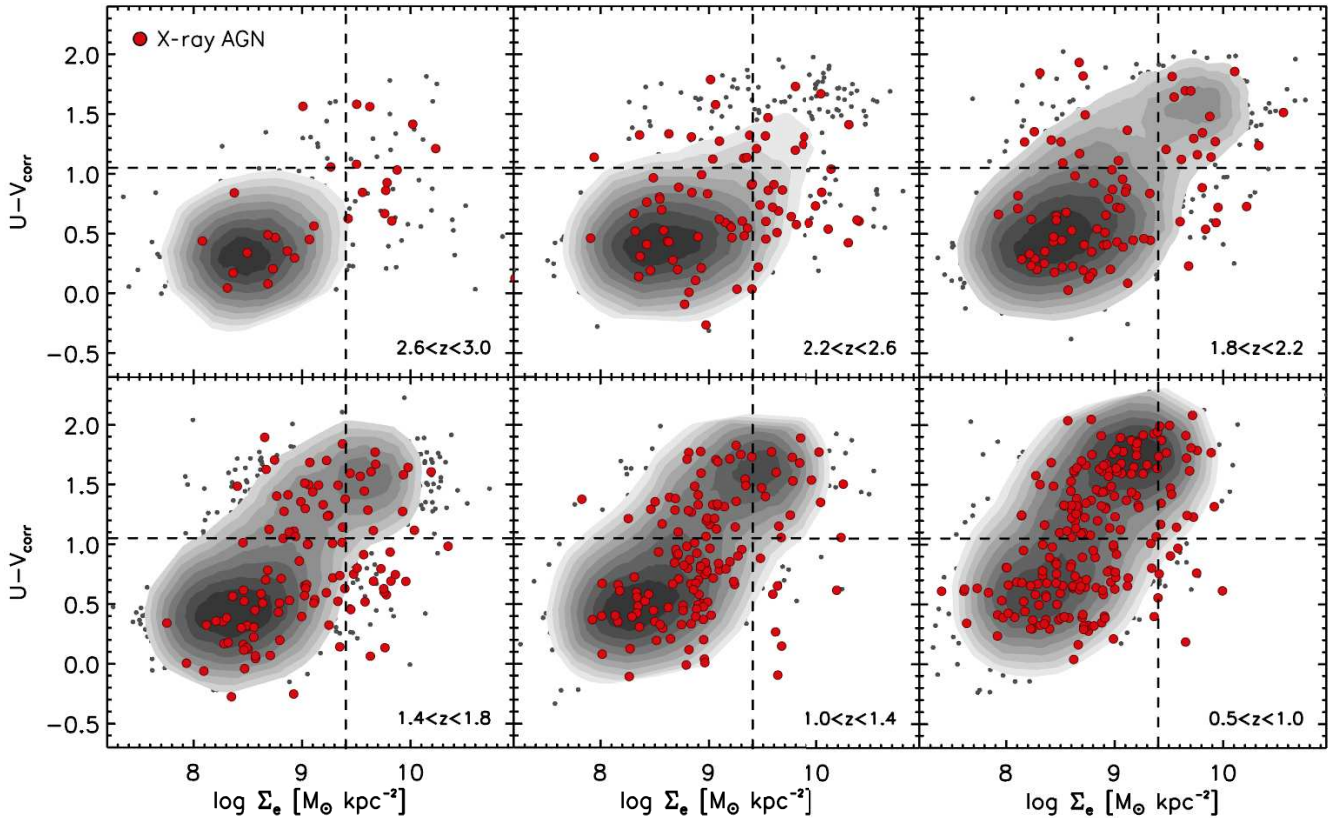}
\caption{Surface mass density ($\Sigma_{e}$) vs rest-frame color for galaxies with $M_{*}>10^{10}$ $M_{\odot}$ in various redshift bins.  Galaxies that host an X-ray AGN are shown as red circles, while non-AGN are denoted with black circles / grey contours.  Compact galaxies lie to the right of the vertical dashed line, while star-forming and passive galaxies are roughly split by the horizontal dashed line.  
  \label{fig-sc_zmosaic}}  
\end{figure*}

Nonetheless, it can be argued that even moderate luminosity AGN may significantly contaminate our color, mass and size measurements.  There are several lines of evidence that suggest this is not the case.  First, efforts to model the SED of AGN hosts in the CANDELS fields using galaxy+AGN hybrid templates have found that the mean color contamination ($\Delta(U-V)$) from non-stellar light for type I and type II AGN is -0.44 and 0.07 magnitudes, respectively (Hsu et al., in prep.).  This suggests that color contamination by type II AGN, which make up the bulk of our sample, is negligible.   Second, Santini et al.~(2012) computed AGN host masses in the GOODS and COSMOS fields by decomposing the total emission of X-ray sources into stellar and nuclear components.  They find that for type II AGN, the relative difference between the stellar mass computed using pure stellar templates and the mass determined using their decomposition technique is consistent with zero.  Furthermore, they report that only 1.3\% of sources had a difference in their stellar mass larger than a factor of two.  Finally, Barro et al.~(2016) recently obtained spatially-resolved ALMA 870 $\mu m$ dust continuum observations of several X-ray detected cSFGs in GOODS-S and confirmed that their compact size is not the result of unresolved nuclear emission.  In fact, the dust continuum emission was found to be twice as compact as the rest-frame optical emission as measured in the WFC3 $H$-band (see also, Ikarashi et al.~2015; Tadaki et al.~2015, 2017).  Based on this body of work, our exclusion of visibly contaminated hosts, and our surface brightness profile tests, we are confident that the galaxy properties that we measure for the remaining AGN hosts are not significantly affected by nuclear emission.

\section{Results}

In Figure \ref{fig-sc_zmosaic}, the surface density, $\Sigma_{e}$, of our parent sample of massive galaxies ($M_{*}>10^{10}$ $M_{\odot}$) is plotted against their dust corrected $U-V$ color in six redshift slices over the range $0.5<z<3.0$.    Star-forming galaxies, as selected by their $UVJ$ colors, typically lie below the horizontal dashed line at $(U-V)_{\rm corr} = 1.08$.  Our surface density threshold for selecting compact galaxies ranges from log $\Sigma_{e}=9.2-9.9$ $M_{\odot}$ kpc$^{2}$, depending on the mass of the galaxy.  To guide the eye, the vertical dashed line denotes log $\Sigma_{e} = 9.4~M_{\odot}$ kpc$^{-2}$; roughly 90\% of our cSFGs have surface densities above this value.  As noted by Barro et al.~(2013), the number density of cSFGs increases from $z=3$ to $z=1.4$ and then rapidly declines at $z<1.4$.   

\begin{figure*}[t]
\epsscale{1.15}
\plotone{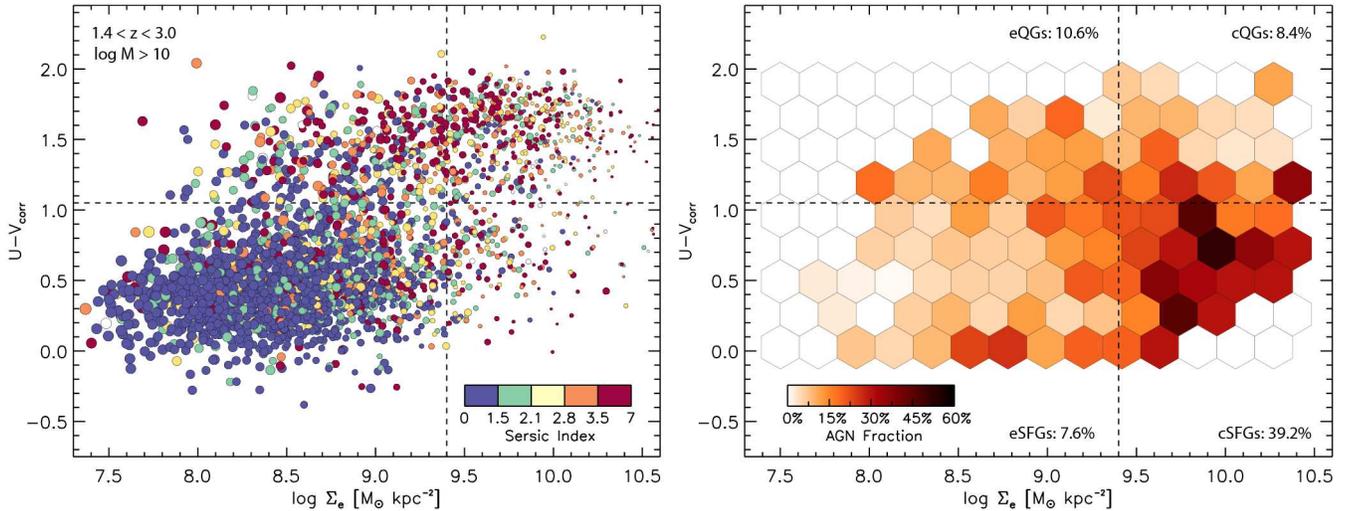}
\caption{({\emph left}) Surface mass density ($\Sigma_{e}$) versus rest-frame color for galaxies with $M_{*}>10^{10}$ $M_{\odot}$ in the redshift range $1.4<z<3.0$.  Points are color coded by their best-fit Sersic index and symbol sizes are scaled to the physical size of each galaxy.  ({\emph right}) AGN fraction in regions of $\Sigma_{e}$-color space.  We find the AGN fraction peaks among the compact, star-forming population.
  \label{fig-hexmap}}  
\end{figure*}

To quantify the AGN fraction among different galaxy populations at $z\sim2$, Figure \ref{fig-hexmap} plots $\Sigma_{e}$ versus dust corrected $U-V$ color over the redshift range where the number density of cSFGs peaks, $1.4<z<3.0$.
Points are again color coded based on their Sersic index and the symbol size is scaled to the physical size of each galaxy.  On the right panel of Figure \ref{fig-hexmap} is shown the AGN fraction in regions of the $\Sigma_{e}$-color space.  Our measured AGN fractions are also listed in Table 1.  

The overall AGN fraction among our parent sample of galaxies with $M_{*}>10^{10}$ $M_{\odot}$ and $1.4<z<3.0$ is $9.7\pm0.5$\% (310 AGN in 3199 galaxies).  Within this sample, there are a total of 166 cSFGs, of which 65 host an X-ray AGN, resulting in an AGN fraction of $39.2^{+3.9}_{-3.6}$\%.   This is significantly higher than the AGN fraction measured among extended star-forming galaxies (eSFGs) in the same mass and redshift range.  There are 2279 eSFGs in our parent sample, of which 174 host an AGN, resulting in an AGN fraction of $7.6^{+0.5}_{-0.6}$\%.  Therefore, cSFGs host X-ray luminous AGN 5.2 times more often relative to their more extended, star-forming counterparts.  This difference is significant at the $8.6\sigma$ level.  

\begin{center}
\tabletypesize{\scriptsize}
\begin{deluxetable}{lccc}
\tablewidth{0pt}
\tablecaption{Fraction of Galaxies Hosting X-ray AGN at $1.4<z<3.0$}
\tablecolumns{4}
\tablehead{\colhead{Sample} & \colhead{$N_{\rm Galaxies}$} & \colhead{$N_{\rm AGN}$} & \colhead{AGN Fraction} }
		   \startdata
		   
Parent Sample\tablenote{\footnotesize Defined as galaxies with $M_{*}>10^{10}$ $M_{\odot}$ and $1.4<z<3.0$}  & 3199  &  310  &  $09.7^{+0.5}_{-0.5}$\%            \\
Compact Star-Forming     &  166  &   65  &  $39.2^{+3.9}_{-3.6}$\%   \\
Extended Star-Forming    & 2279  &   174 &  $07.6^{+0.5}_{-0.6}$\%    \\
Compact Quiescent        &  404  &    34 &  $08.4^{+1.2}_{-1.6}$\%    \\
Extended Quiescent       &  350  &    37 &  $10.6^{+1.7}_{-1.6}$\%   \\
\hline
Mass-Matched             &       &       &                          \\
Extended Star-Forming    &  308  &    38 &  $12.3^{+1.6}_{-2.1}$\%    \\
\vspace*{-0.12in}
\enddata

\end{deluxetable}
\end{center} 

\vspace*{-0.35in}

However, the X-ray AGN fraction is known to increase with galaxy mass (e.g., Xue et al.~2010; Aird et al.~2012).  If the cSFGs are systematically more massive than their eSFGs counterparts, this may explain their elevated AGN fraction.  In fact, this appears to be the case.  The median mass of the cSFGs is log $<$$M_{*}$$>\ = 10.74$ $M_{\odot}$, while that of the eSFGs is log $<$$M_{*}$$>\ = 10.19$ $M_{\odot}$.  To account for this, we have constructed a mass-matched sample of eSFGs.  For every cSFGs, we randomly selected two eSFGs whose mass is within a factor of two and redshift within $\Delta z = 0.2$ of the cSFG.  Using this selection, the median masses of the two populations are in much better agreement with log $<$$M_{*}$$>\ = 10.74$ $M_{\odot}$ and 10.70 $M_{\odot}$ for the compact and extended systems, respectively.

Matching the mass distribution of the eSFGs to that of the cSFGs results in an increase in their AGN fraction.  We find that $12.3^{+1.6}_{-2.1}\%$ of the mass-matched eSFGs host an X-ray AGN.  Despite this increase, the AGN fraction in cSFGs is still higher than the fraction measured in the eSFGs of similar mass.  After controlling for mass, cSFGs host AGN 3.2 times more often than eSFGs, a difference that is significant at the $6.2\sigma$ level.  

The AGN fraction in cSFGs is also elevated relative to the quiescent galaxy population. We find that $8.4^{+1.2}_{-1.6}$\% of compact quiescent galaxies (cQGs) host an AGN (34 out of 404 galaxies), while the same is true for $10.6^{+1.7}_{-1.6}$\% of extended quiescent galaxies (eQGs), where we find 37 AGN in 350 galaxies.  These fractions are 4.7 and 3.7 times smaller ($8.0\sigma$ and $7.0\sigma$ differences) than the fraction observed among the cSFGs, respectively.  Overall, we find that among the massive galaxy population at $z>1.4$, X-ray AGN are most prevalent in compact, star-forming systems.

Of course, the AGN fractions reported in this section are only lower limits and subject to the flux limits of the existing X-ray data.  Deeper X-ray observations in the UDS and EGS fields, for example, would certainly increase these fraction as additional AGN with lower luminosities are detected.  If we limit our analysis to GOODS-South, which has the deepest \emph{Chandra} data of our four fields, we can construct a volume-limited sample of AGN with X-ray luminosities of $L_{\rm 0.5-8 keV}>5\times10^{42}$ erg s$^{-1}$ out to $z=3$.  Using this sample, we find that $55.9^{+8.6}_{-7.9}$\% of cSFGs with masses of $M_{*}>10^{10}$ $M_{\odot}$ in the redshift range $1.4<z<3.0$ host an X-ray bright AGN.  This fraction agrees with the results of Barro et al.~(2014), who reported that roughly half of their cSFGs in GOODS-South are X-ray detected.

Based on number density arguments and gas depletion timescales, the lifetime of the compact, star-forming phase is estimated to be roughly $\sim$500 Myr (van Dokkum et al.~2015; Barro et al.~2016).  Therefore, the AGN fraction we measure in our volume-limited sample implies a duty cycle as long as $\sim280$ Myr.  This is consistent with, although on the higher end, of AGN duty cycles reported in the literature, which typically range from tens to hundreds of Myr (e.g., Haehnelt et al.~1998; Mathur et al.~2001; Shabala et al.~2008).  

Finally, we note that while our X-ray luminosity limit for selecting AGN, $L_{\rm X}>10^{42}$ erg s$^{-1}$, is lower than the canonical limit of $3.2\times10^{42}$ erg s$^{-1}$ (e.g., Padovani et al.~2017), using this higher selection threshold in all four of our fields does not significantly affect our results.  We find that $38.6^{+3.9}_{-3.6}$\% of cSFGs host an AGN with $L_{\rm X} > 3.2\times10^{42}$ erg s$^{-1}$ versus only $6.7^{+0.5}_{-0.6}$\% of eSFGs in the redshift range $1.4<z<3.0$.  When the eSFG sample is matched in mass to the cSFGs, their AGN fraction increases to $12.9^{+1.6}_{-2.1}$\%.   In summary, the enhancement of AGN activity in the cSFG population remains even when a more conservative X-ray luminosity threshold is employed.

\section{Discussion}

\subsection{Triggering Mechanisms}

The increased AGN fraction that we find in cSFGs implies either an increase in the AGN duty cycle among this population or an increase in their accretion efficiency.  In either case, this suggests the same physical processes that give rise to the compact star formation activity in these galaxies may also aid in funneling gas to their centers, thereby triggering the increased AGN activity we observe.  An evolutionary pathway has been proposed in which cSFGs are the descendants of larger, more extended star-forming galaxies that underwent a compaction phase as a result of gas-rich, dissipational processes, such as wet mergers or violent disk instabilities (Barro et al.~2013).  In this scenario, one or more nuclear starbursts drive a rapid increase in the galaxy's central stellar density and a decrease in its half-mass radius (Dekel et al.~2009; Dekel \& Burkert 2014).

\begin{figure*}[t]
\epsscale{1.15}
\plotone{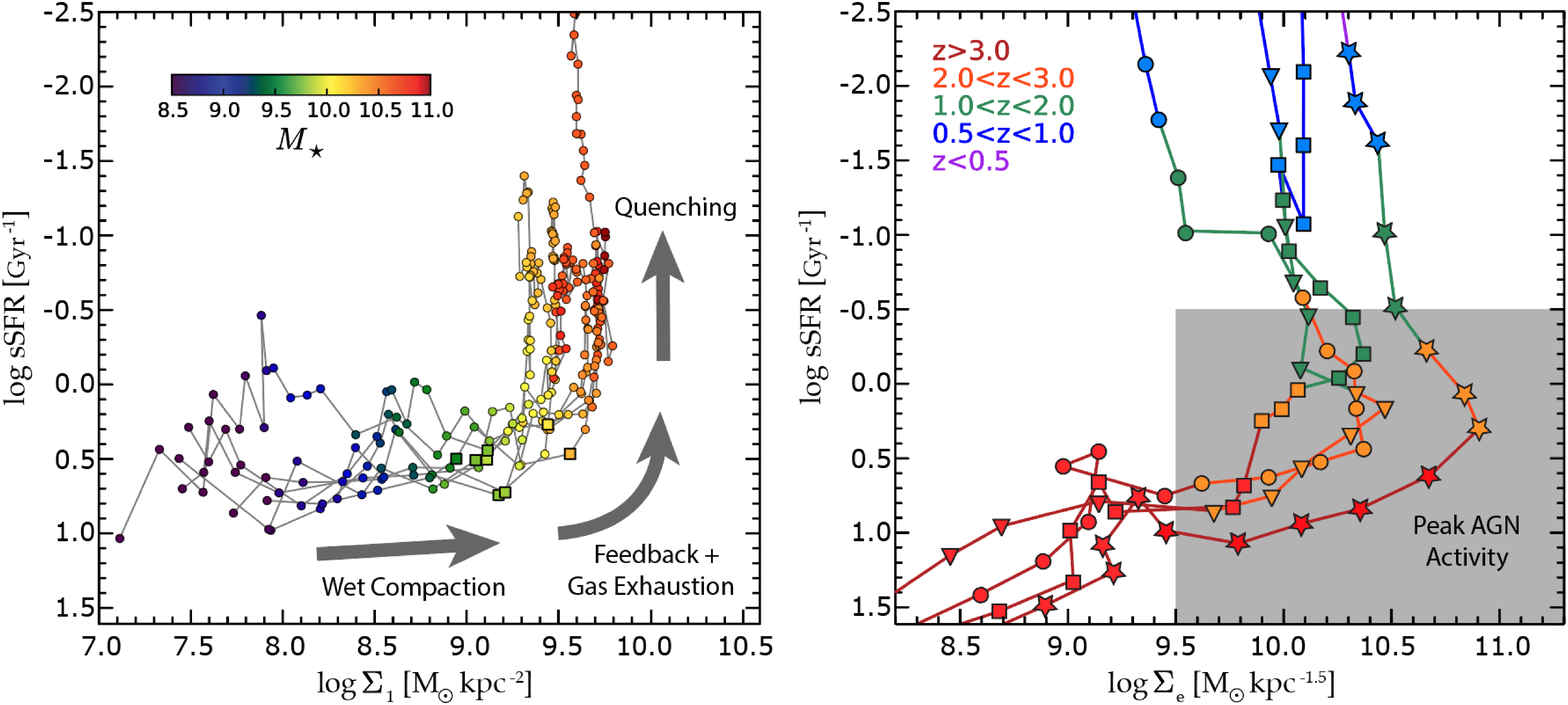}
\caption{({\emph left}) Predicted evolution in surface mass density, $\Sigma$, versus specific SFR of eight galaxies that experience a wet compaction phase in the VELA simulations.  Here surface density is measured within the central 1 kpc ($\Sigma_{\rm 1 kpc}$, see Appendix) and the color coding specifies the system's stellar mass.  Squares pinpoint when maximum gas compaction is achieved in each galaxy.  The compaction results in a nuclear starburst that rapidly increases the simulated galaxies' central stellar density and decreases their half-mass radii.  Eventually their star formation is quenched due to a combination of gas exhaustion and feedback processes, resulting in a massive, compact remnant.  We find that AGN activity peaks in the compact, star-forming population; precisely where feedback energy is needed to facilitate quenching.  (\emph{right}) Evolutionary tracks for four galaxies that undergo a wet compaction phase in the Santa Cruz semi-analytics models.  Here surface density is measured within the effective radius and the color coding specifies the galaxy's redshift.  Our results indicate the AGN fraction peaks in compact galaxies that are on the verge of quenching.
  \label{fig-theory}}  
\end{figure*}

Indeed, cosmological zoom-in simulations show that dissipative contraction triggered by intense gas inflow episodes at $z\sim2-4$ can produce galaxies with similar surface mass densities as the observed cSFG population (Zolotov et al.~2015; Tacchella et al.~2016a).  Figure \ref{fig-theory} shows evolutionary tracks from the cosmological hydro-dynamic VELA simulations that follow the structural evolution of massive galaxies that undergo a wet compaction phase (Dekel et al., in prep.).  Also shown are similar tracks from the Santa Cruz semi-analytic models (Somerville \& Primack 1999; Somerville et al.~2008; Porter et al.~2014).  In both cases, a dissipative contraction results in a nuclear starburst which rapidly increases the central stellar density of the galaxies.  Furthermore, results from the Illustris simulation (not shown) predict that the same high gas densities that give rise to the nuclear starburst will also fuel concurrent AGN activity (Wellons et al.~2015), in excellent agreement with our findings (see also Habouzit et al., in prep.).

This framework is supported observationally by spatially-resolved ALMA observations that have revealed intense star formation activity (SFR$=200-700$ $M_{\odot}$ yr$^{-1}$) on nuclear scales ($r_{e} \sim 1$ kpc) in cSFGs at $z\sim2$ (Barro et al.~2016; Tadaki et al.~2017).  This activity is estimated to increase the stellar mass density of these galaxies by $4\times$ and decrease their half-mass radii by $1.6\times$ on relatively short timescales ($t\sim200$ Myr).  

If this formation scenario is correct, then the elevated AGN activity that we find indicates the inflow episodes which produce the wet compaction phase are also effective at triggering increased SMBH growth in these galaxies.  In fact, the compaction and rapid mass build-up may help overcome the effects of supernova feedback in suppressing accretion on to the central black hole at lower masses pre-compaction, especially once the halo potential well becomes deeper than the critical value for supernova feedback ($V>100$ km s$^{-1}$; Dekel \& Silk 1986), corresponding to $M_{\rm vir}\sim10^{12}$ $M_{\odot}$.  At such masses, supernova feedback becomes inefficient, resulting in gravitationally bound gas that can continue accreting onto the SMBH and fueling subsequent AGN activity.

Our observations are consistent with previous findings that the AGN fraction increases substantially in blue bulge-dominated galaxies (e.g., Silverman et al.~2008) and those undergoing periods of intense star formation activity (Kartaltepe et al.~2010; Xue et al.~2010; Juneau et al.~2013).  An alternative formation scenario for cSFGs has proposed that their dense cores formed in situ at even higher redshifts, when all galaxies were denser (Lilly \& Carollo 2016; Williams et al.~2017).  However, this scenario predicts a concurrent phase of elevated SMBH growth at higher redshifts ($z=3-5$; Wellons et al.~2015), which appears to be at odds with the increased AGN activity we find down to $z\sim1.4$.

\subsection{Progenitors of Compact Quiescent Galaxies}

cQGs are among the first generation of massive, passively evolving galaxies to appear in large numbers at $z\sim2-3$.  Several studies have shown that the number density of cQGs increases between $z\sim3$ and $z\sim1.4$ at the expense of cSFGs, which become increasingly rare at $z<2$ (Barro et al.~2013, 2014; van Dokkum et al.~2015).  As a result, Barro et al.~(2013) proposed an evolutionary link between these two populations in which cSFGs are the direct progenitors of cQGs.  With their small sizes, steep mass profiles, and similar kinematic properties, cSFGs are nearly identical to cQGs, implying that only their star formation activity need be quenched for them to passively evolve into their quiescent counterparts.  Based on the observed increase in the number density of cQGs with time, the quenching timescale for cSFGs is estimated to be relatively short ($\sim500$ Myr; van Dokkum et al.~2015).   

Therefore, the increased AGN fraction that we measure in the compact, star-forming population has two important implications.  First, it indicates that cQGs experience a phase of elevated SMBH growth directly preceding the shut down of their star formation.  It has been estimated that roughly 1/3 of the stellar mass in cQGs is formed during the lifetime of cSFGs (van Dokkum et al.~2015), therefore a period of elevated black hole growth might be expected if these galaxies are to remain on the $M_{\rm BH}-M_{*}$ relationship, which has been observed to already be in place at $z\sim2$ (Mullaney et al.~2012).  

Using similar arguments as van Dokkum et al.~(2015), we can estimate the fraction of black hole mass that is accreted during the cSFG phase.  We start with the mass accretion rate of a SMBH, which is related to the bolometric energy output of an AGN as

\begin{equation}
\dot{M}_{\rm BH} = \frac{L_{Bol}}{c^{2}\eta}
\end{equation}

\noindent where $\eta$ is the matter to radiation conversion efficiency, which has a expected mean value of $\eta=0.1$ (Thorne 1974, Elvis et al.~2002).  Following Bluck et al.~(2012), we adopt the minimum bolometric correction reported in the literature, 15.0, to convert $L_{\rm X}$ (0.5-8 kev) to $L_{\rm Bol}$, which gives a median bolometric luminosity of $L_{\rm Bol}=5.76\times10^{44}$ erg s$^{-1}$ for our sample of cSFGs.  This results in a minimum accretion rate of $\dot{M}_{\rm BH} = 0.1$ $M_{\odot}$ yr$^{-1}$.  With a duty cycle of 280 Myr (see \S4), a total mass of $M_{\rm BH,cSFG}=2.8\times10^{7}$ $M_{\odot}$ is accreted during the compact star-forming phase before these galaxies quench.   

If we assume that cSFGs follow the local $M_{\rm BH}-M_{*}$ relationship, then we can determine what fraction of their SMBH mass this newly accreted mass represents.  Using the relation of Haring \& Rix (2004), the stellar mass of our cSFG sample implies a median SMBH mass of $M_{\rm BH} = 7.7\times10^{7}$ $M_{\odot}$.  Following van Dokkum et al.~(2015), we assume that the cSFGs are observed, on average, halfway through their lifetimes, therefore their final SMBH mass before quenching will be

\vspace*{-0.1in}
\begin{equation}
 M_{\rm BH, final} = M_{\rm BH} + 0.5 M_{\rm BH,cSFG} = 9.1\times10^{7} M_{\odot}.  
\end{equation}

\noindent This means that roughly 31\% ($2.8\times10^{7}/9.1\times10^{7}$) of the SMBH mass contained in cQGs is accreted during the cSFG phase. This is in good agreement with the estimated fraction of stellar mass formed during this period ($\sim1/3$) as reported by van Dokkum et al.~(2015). This suggests that even with moderate X-ray luminosities, the elevated AGN activity that we observe in cSFGs may be key to maintaining/establishing the $M_{\rm BH}-M_{*}$ relationship in their quiescent descendants.

The second implication of our findings is that the increased AGN activity in cSFGs raises the possibility that feedback from the AGN may play a role in quenching their star formation activity.  Indeed, galaxy evolution simulations and models indicate a substantial amount of energy injection is needed in the cSFG phase in order to achieve quenching timescales that are consistent with the number density evolution of cQGs.   In the VELA tracks shown in Figure \ref{fig-theory}, the starburst activity in cSFGs is eventually quenched through a combination of gas exhaustion and feedback processes, which results in their relatively rapid migration onto the red sequence.  While the VELA simulations do not currently include AGN feedback, Zolotov et al.~(2015) note that without the additional energy injection from sources such as AGN, full quenching to very low specific SFRs does not fully occur in the timescale needed to ensure the absence of cSFGs by $z\sim1.4$ (see also Pandya et al.~2017; Brennan et al.~2017).  This means that the observed AGN fraction peaks in precisely the population where simulations predict feedback energy is vital in order to reproduce the number density evolution of cQGs at $z\sim2$.

That said, it is still debated whether energy from an AGN is necessarily needed to quench cSFGs.  Using ALMA observations, Barro et al.~(2016) find gas depletion timescale of $M_{\rm gas}/$SFR$=230$ Myr, implying the starburst activity of these galaxies may be short lived assuming no further gas is accreted onto the system.  Recent results with VLA and ALMA further confirm the short depletion times of these galaxies (Spilker et al.~2016; Popping et al.~2017).  In addition, a population of compact starburst galaxies found at $z\sim0.6$ by Diamond-Stanic et al.~(2012), which may be rare analogs of cSFGs at lower redshifts, exhibit high velocity outflows ($>1000$ km s$^{-1}$) in the apparent absence of any concurrent nuclear activity.   It has been proposed that these extreme velocities are related to their compact starburst activity, which can deposit a large amount of momentum in an unusually small region (Heckman et al.~2011).   However, the cSFG population at $z\sim2$ differs from their rare, lower redshift counterparts in that their AGN activity appears to be ubiquitous.

Even in the absence of ejective feedback driven by the AGN (Hopkins et al.~2008), the elevated nuclear activity that we observe may help prevent future star formation in these galaxies.  By rapidly building up the central black hole, the compact phase may help halt further accretion through heating/outflows driven by radiatively inefficient relativistic jets (e.g., Croton et al.~2006; Choi et al.~2015).  In this scenario, cSFGs would initially quench due to gas exhaustion and non-AGN feedback processes and thereafter remain quenched due to preventative feedback from the SMBH.  Future simulations that include AGN feedback should help determine which of these mechanisms, ejective or preventative feedback, is more important (or perhaps both are equally important) in ultimately quenching cSFGs.  Further observational work is also needed to determine if these systems are experiencing large scale outflows that may aid in quenching their star formation activity and whether the elevated nuclear activity that we observe plays a role in driving these winds.

\section{Conclusions}

We have examined the prevalence of AGN activity in massive ($M>10^{10} M_{\odot}$), compact star-forming galaxies at $1.4<z<3$ in four of the CANDELS fields using deep X-ray observations.  These galaxies are likely the direct progenitors of compact, quiescent galaxies, which are the first population of passive galaxies to appear in large numbers at $z=2-3$.  We select compact galaxies using a mass-dependent surface mass density, $\Sigma_{e}$, threshold that is 0.3 dex below the $\Sigma_{e}$-mass relationship for quiescent galaxies at $z\sim2$.  Star-forming and quiescent systems are distinguished by their rest-frame $UVJ$ colors and AGN hosts are identified as galaxies with X-ray luminosities of $L_{\rm X} > 10^{42}$ erg s$^{-1}$ as measured in \emph{Chandra} imaging of the fields.

We take care to remove AGN hosts contaminated by nuclear light from our analysis by excluding unresolved hosts and extended hosts that show point-like emission at their centers.  Such sources are identified using a combination of visual inspection and surface brightness profile fitting.  Our Galfit modeling and stacked surface brightness profiles indicate that, on average, less than 10\% of the rest-frame optical light from the X-ray detected cSFGs originates from an unresolved nuclear component.

Based on a sample of 3199 massive galaxies ($M>10^{10} M_{\odot}$) in the redshift range $1.4<z<3$, our primary results are as follows:

\begin{itemize}
\item[1.] We find that among galaxies with $M>10^{10} M_{\odot}$, X-ray luminous AGN activity is most prevalent in cSFGs; $39.2^{+3.9}_{-3.6}$\% of such galaxies host an AGN, compared to only $7.6^{+0.5}_{-0.6}$\% of larger eSFGs.  This $5.2\times$ difference is significant at the $8.6\sigma$ level.

\item[2.] Using a mass-matched sample of eSFGs reduces this disparity, but does not eliminate it.  We find that $12.3^{+1.6}_{-2.1}\%$ of the mass-matched eSFGs host an X-ray AGN; a decrement of $3.2\times$ relative to cSFGs that is significant at the $6.2\sigma$ level.

\item[3.]  cSFGs also host AGN more often than compact and extended quiescent galaxies with $M>10^{10} M_{\odot}$.  We find that $8.4^{+1.2}_{-1.6}$\% of cQGs and $10.6^{+1.7}_{-1.6}$\% of eQGs host an X-ray detected AGN, which is 4.7 and 3.6 times less often than cSFGs.  These differences are significant at the $8.0\sigma$ and $7.0\sigma$ levels, respectively.
  
\item[4.] Using a volume-limited sample of AGN in the GOODS-South field with $L_{\rm 0.5-8 keV}>5\times10^{42}$ erg s$^{-1}$ and $1.4<z<3$, we find that $55.9^{+8.6}_{-7.9}$\% of cSFGs host an AGN.  Based on the expected lifetime of cSFGs ($\sim$500 Myr), this fraction implies a AGN duty cycle as long as $\sim280$ Myr during this compact phase.
  
\end{itemize}

Our findings suggest that the same physical mechanisms that trigger the intense star formation observed in cSFGs are also effective at fueling contemporaneous SMBH growth.  Our results are in general agreement with formation scenarios in which cSFGs are created by the dissipative contraction of gas-rich galaxies that triggers in a nuclear starburst and elevated AGN activity.  Although we cannot directly test formation scenarios that propose cSFGs are relics of an earlier formation epoch (when all galaxies have smaller sizes), these models favor elevated AGN activity at higher redshifts ($z>3$) than that reported here.

Given that cSFGs are expected to quench on timescales of $\sim$500 Myr, our results indicate that their quiescent descendants, cQGs, experience a phase of elevated SMBH growth directly preceding the shutdown of their star formation.  Based on their X-ray luminosities and the duty cycle implied by our AGN fractions, we estimate that $\sim$31\% of the SMBH mass contained in cQGs is accreted during the cSFG phase.  This roughly matches the estimated 1/3 of stellar mass formed during this phase, suggesting this period of growth is key to maintaining/establishing the $M_{\rm BH}-M_{*}$ relationship in galaxies that are the likely progenitors of today's giant ellipticals.

Finally, we note that the increased AGN activity that we observe in cSFGs may be related to their imminent quenching, as galaxy evolution simulations and models require feedback energy during this phase in order to reproduce the short quenching timescale reported in the literature.  However, further work is needed to determine if these galaxies are experiencing large-scale outflows and what role AGN may play in driving this activity.

\vspace{0.25in}
Support for Program number HST-GO-12060 was provided by NASA through a grant from the Space Telescope Science Institute, which is operated by the Association of Universities for Research in Astronomy, Incorporated, under NASA contract NAS5-26555.

\appendix
\section{\bf AGN Fraction versus Central Surface Mass Density, $\Sigma_{\rm 1~kpc}$} 

Throughout this paper, we have identified cSFGs based on their surface mass density measured at the effective radius, $\Sigma_{e}$.  However, it has recently been argued that a more robust indicator of compactness is instead surface mass density measured within the central 1 kpc, $\Sigma_{\rm 1~kpc} = 0.5M_{*}({\rm <1~kpc})/\pi ({\rm 1~kpc})^{2}$ (e.g., Barro et al.~2017).  Several studies have shown that central core density is more tightly correlated with stellar mass than the effective radius and more closely related with quiescence (Cheung et al.~2012; van Dokkum et al.~2014; Tacchella et al.~2015; Whitaker et al.~2017).  Therefore, in this section, we demonstrate that our primary result, an enhancement of AGN activity in cSFGs, holds true if we use $\Sigma_{\rm 1~kpc}$ to select compact galaxies instead of $\Sigma_{e}$.  For this test, we limit our analysis to galaxies brighter than $H=24.5$ mag, more massive than $M_{*}>10^{10} M_{\odot}$, and those at $0.5<z<2.5$ (the redshift range over which we currently have $\Sigma_{\rm 1~kpc}$ measurements).  Our HST/WFC3 \emph{H}-band imaging has a spatial resolution of $r=0.5-0.7$ kpc over this redshift range and is therefore able to resolve the inner 1 kpc of our target galaxies.  As in our primary analysis, star-forming and quiescent galaxies are selected using a $U-V$ and $V-J$ color cut based on that of Williams et al.~(2009).

\begin{figure*}[t]
\epsscale{1.15}
\plotone{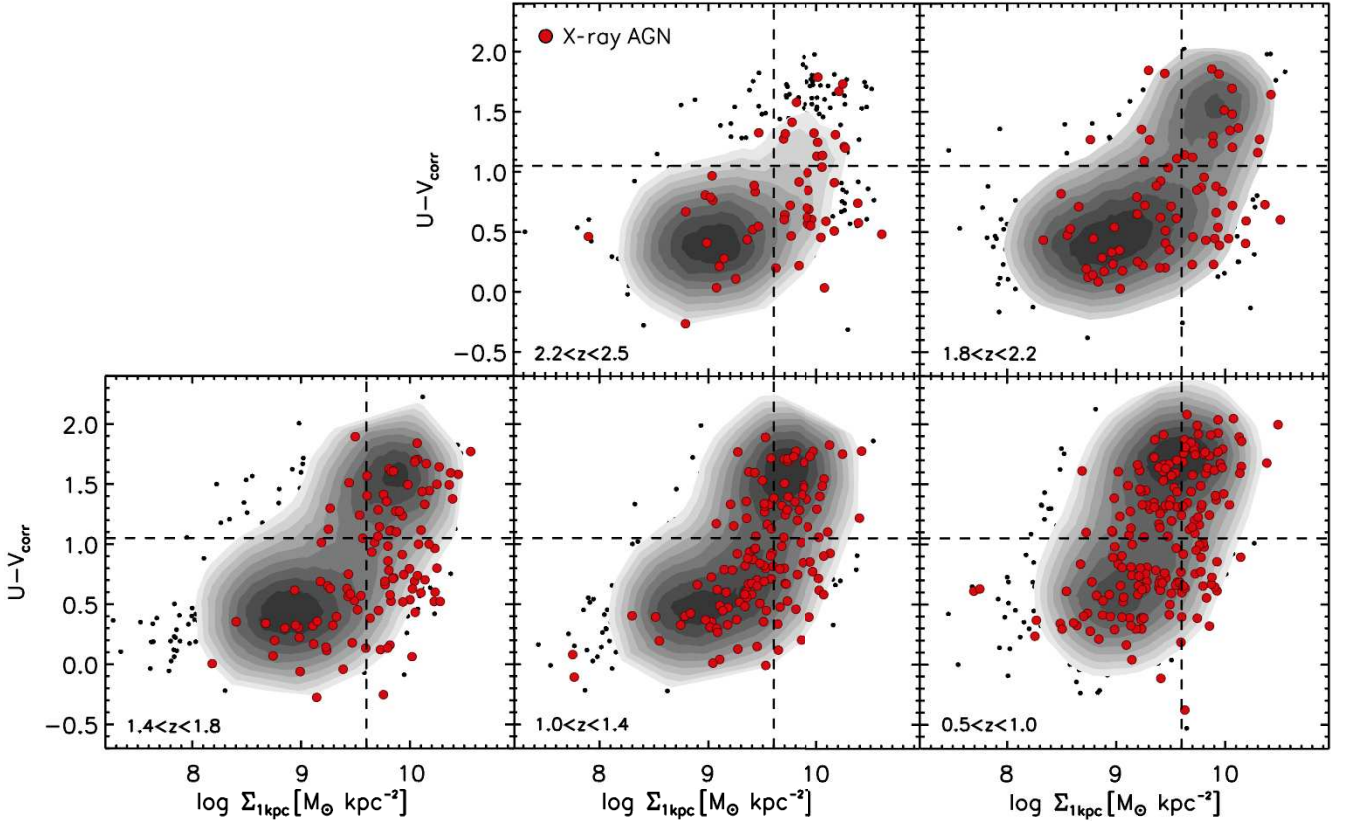}
\caption{Central surface mass density ($\Sigma_{\rm 1~kpc}$) vs rest-frame color for galaxies with $M_{*}>10^{10}$ $M_{\odot}$ in various redshift bins.  Galaxies that host an X-ray AGN are shown as red circles, while non-AGN are denoted with black circles / grey contours.  Compact galaxies lie to the right of the vertical dashed line, while star-forming and passive galaxies are roughly split by the horizontal dashed line.  We find $\Sigma_{\rm 1~kpc}$ to be better correlated with quiescence compared to surface mass density measured at the effective radius, $\Sigma_{e}$.
  \label{fig-sc_zmosaic_sigma1}}  
\end{figure*}

In Figure \ref{fig-sc_zmosaic_sigma1}, the central surface density, $\Sigma_{\rm 1~kpc}$, of our sample of massive galaxies ($M_{*}>10^{10}$ $M_{\odot}$) is plotted against their dust corrected $U-V$ color in five redshift slices over the range $0.5<z<2.5$.  This plot shares many qualitative similarities with Figure \ref{fig-sc_zmosaic}, which plots surface density at the effective radius, $\Sigma_{e}$, versus color.  In both cases, the number density of cSFGs increases to $z=1.4$ and then rapidly declines at lower redshifts.  However, using $\Sigma_{\rm 1~kpc}$ causes a reduction in the number of galaxies in the extended quiescent region of Figure \ref{fig-sc_zmosaic_sigma1} (the upper left quadrant).  This causes a more pronounced transition from extended star-forming systems to compact quiescent galaxies at high values to $\Sigma_{\rm 1~kpc}$, which is consistent with findings that core density is closely related to quiescence (e.g., Cheung et al.~2012; Whitaker et al.~2017).  In other words, quiescent galaxies have a reduced spread in $\Sigma_{\rm 1~kpc}$ versus $\Sigma_{e}$ at any given redshift.  This is likely due to the fact that the effective radius of quiescent galaxies can increase post-quenching due to dry mergers or the re-accretion of material, whereas these processes have a less dramatic impact a galaxy's core density (Nabb et al.~2009; Oser et al.~2010).

\begin{figure*}[t]
\epsscale{1.15}
\plotone{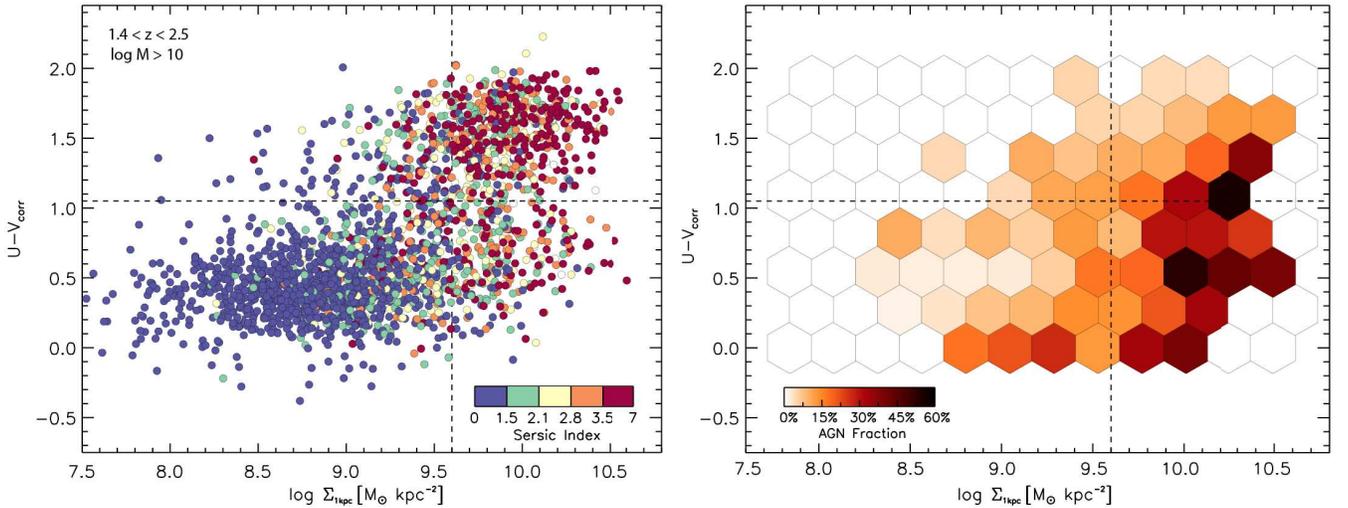}
\caption{({\emph left}) Central surface mass density ($\Sigma_{\rm 1~kpc}$) versus rest-frame color for galaxies with $M_{*}>10^{10}$ $M_{\odot}$ in the redshift range $1.4<z<2.5$.  Points are color coded by their best-fit Sersic index.  ({\emph right}) AGN fraction in regions of $\Sigma_{\rm 1~kpc}$-color space.  We find the AGN fraction peaks among the compact, star-forming population; a result that holds true whether compact galaxies are defined using $\Sigma_{e}$ or $\Sigma_{\rm 1~kpc}$.
  \label{fig-hexmap_sigma1}}
\end{figure*}

The AGN fraction in the $\Sigma_{\rm 1~kpc}$-color plane can be seen in Figure \ref{fig-hexmap_sigma1}, which plots $\Sigma_{\rm 1~kpc}$ versus dust corrected $U-V$ color over the redshift range $1.4<z<2.5$.  We again find that the AGN fraction peaks in star-forming, compact galaxies located in the lower right quadrant of the plot.  This is consistent with our findings using $\Sigma_{e}$ in \S4 and in excellent agreement with the evolutionary tracks from the VELA simulations shown in Figure \ref{fig-theory}, which require feedback energy in the cSFGs phase that directly precedes quenching.  

\begin{figure}[t]
\epsscale{0.9}
\plotone{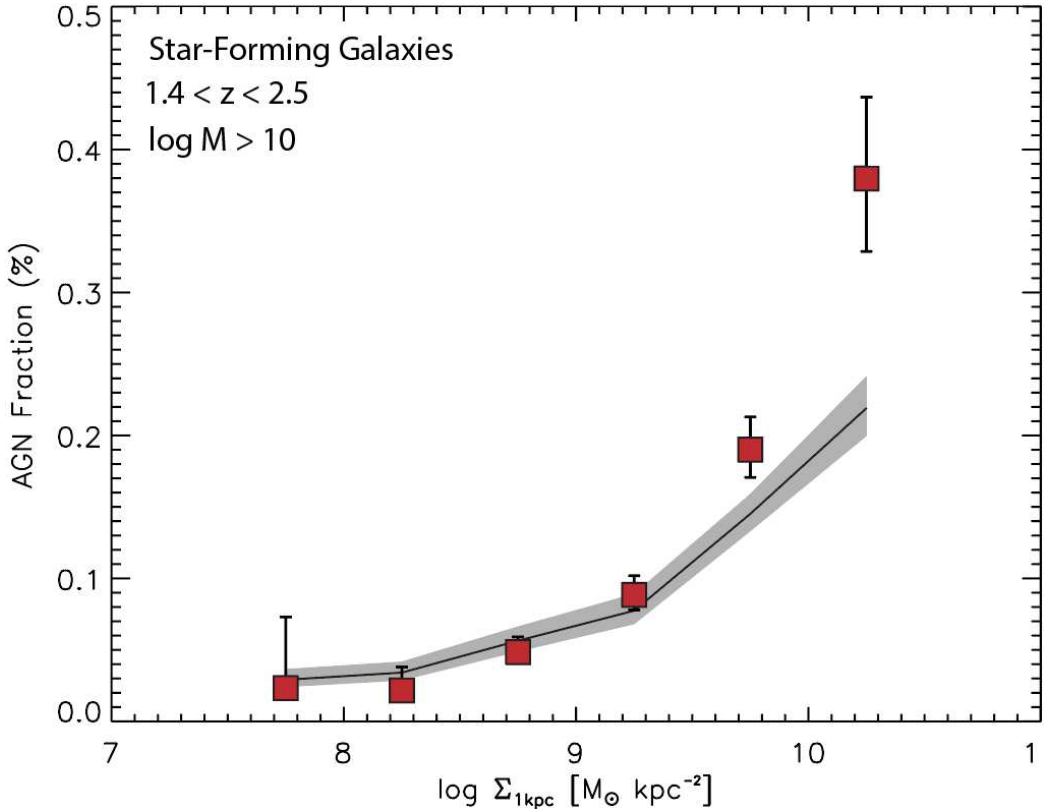}
\caption{AGN fraction versus $\Sigma_{\rm 1~kpc}$ for star-forming galaxies with $M_{*}>10^{10}$ $M_{\odot}$ in the redshift range $1.4<z<2.5$.  The black solid line shows the expected AGN fraction in galaxies matched in mass to those in the $\Sigma_{\rm 1~kpc}$ bins.  Error bars and the grey shaded region show the 68.3\% binomial confidence limits.  At the highest values of $\Sigma_{\rm 1~kpc}$, we find the AGN fraction exceeds what we expect in a mass-matched control sample, suggesting the increased nuclear activity in these galaxies is linked to their high core density.
  \label{fig-agnfrac_sigma1}}
\end{figure}

In fact, we find a strong correlation between AGN activity and core density, as can be seen in Figure \ref{fig-agnfrac_sigma1}.  The dramatic increase of the AGN fraction with $\Sigma_{\rm 1~kpc}$ is partially a mass-driven effect: galaxies with higher core densities are systematically more massive and the X-ray AGN fraction is known to increase with galaxy mass.  However, when we account for this mass difference, we still find an elevated AGN fraction in star-forming galaxies with the highest core densities.  The grey line in Figure \ref{fig-agnfrac_sigma1} shows the AGN fraction in star-forming galaxies which were selected to have masses similar to the galaxies in each $\Sigma_{\rm 1~kpc}$ bin, but irrespective of their core density.  At the highest values of $\Sigma_{\rm 1~kpc}$, we find the AGN fraction exceeds what we expect from the mass-dependance alone.  This suggests the increased nuclear activity in these galaxies is linked to their high core density.  This is in agreement with our results in \S4, where we found elevated AGN activity in cSFGs relative to eSFGs even when their mass difference is taken into account.

Overall, we find that AGN activity at $z\sim2$ peaks among massive, cSFGs that appear to be on the verge of quenching.  Although  $\Sigma_{\rm 1~kpc}$ is better correlated with quiescence than $\Sigma_{e}$, we find that our results hold regardless of whether the surface mass density is measured at the effective radius or in the inner 1 kpc of our target galaxies.

\bibliography{ms.astroph.bbl}

\end{document}